\documentclass[10pt,aps,prl,twocolumn,superscriptaddress,floatfix,nofootinbib]{revtex4-1}

\usepackage{dcolumn}
\usepackage{bm}
\usepackage{amsmath,amssymb,mathrsfs}
\usepackage{ascmac}

\usepackage{geometry}
\usepackage{graphicx}
\def\0\\{\nonumber\\}
\def\bs#1{\boldsymbol{#1}}

\def\br{\mathbf{r}}
\def\zI{\mathrm{i}\hspace{0.2mm}}

\begin{document}

\title{Exploring Zeptosecond Quantum Equilibration Dynamics: \\From Deep-Inelastic to Fusion-Fission Outcomes in $^{58}$Ni+$^{60}$Ni Reactions}
\author{E. Williams}
\email{elizabeth.williams@anu.edu.au}
\affiliation{Department of Nuclear Physics, Research School of Physical Sciences and Engineering, The Australian National University, Canberra, ACT 2601, Australia}

\author{K. Sekizawa}
\altaffiliation{Current address: Department of Physics, University of Washington, Seattle, WA 98195-1560, USA}
\affiliation{Faculty of Physics, Warsaw University of Technology, Ulica Koszykowa 75, 00-662 Warsaw, Poland}

\author{D. J. Hinde}
\affiliation{Department of Nuclear Physics, Research School of Physical Sciences and Engineering, The Australian National University, Canberra, ACT 2601, Australia}

\author{C. Simenel}
\affiliation{Department of Nuclear Physics, Research School of Physical Sciences and Engineering, The Australian National University, Canberra, ACT 2601, Australia}

\author{M. Dasgupta}
\affiliation{Department of Nuclear Physics, Research School of Physical Sciences and Engineering, The Australian National University, Canberra, ACT 2601, Australia}

\author{I. P. Carter}
\affiliation{Department of Nuclear Physics, Research School of Physical Sciences and Engineering, The Australian National University, Canberra, ACT 2601, Australia}

\author{K. J. Cook}
\affiliation{Department of Nuclear Physics, Research School of Physical Sciences and Engineering, The Australian National University, Canberra, ACT 2601, Australia}

\author{D. Y. Jeung}
\affiliation{Department of Nuclear Physics, Research School of Physical Sciences and Engineering, The Australian National University, Canberra, ACT 2601, Australia}

\author{S. D. McNeil}
\affiliation{Department of Nuclear Physics, Research School of Physical Sciences and Engineering, The Australian National University, Canberra, ACT 2601, Australia}

\author{C. S. Palshetkar}
\altaffiliation{Current address: Department of Nuclear and Atomic Physics, Tata Institute of Fundamental Research, Mumbai 400 005 India}
\affiliation{Department of Nuclear Physics, Research School of Physical Sciences and Engineering, The Australian National University, Canberra, ACT 2601, Australia}

\author{D. C. Rafferty}
\affiliation{Department of Nuclear Physics, Research School of Physical Sciences and Engineering, The Australian National University, Canberra, ACT 2601, Australia}

\author{K. Ramachandran}
\altaffiliation{Current address: Nuclear Physics Division, Bhabha Atomic Research Centre, Mumbai 400 085, India}
\affiliation{Department of Nuclear Physics, Research School of Physical Sciences and Engineering, The Australian National University, Canberra, ACT 2601, Australia}

\author{A. Wakhle}
%\altaffiliation{Current address: National Superconducting Cyclotron Laboratory, Michigan State University, East Lansing, MI 48824 USA}
\affiliation{Department of Nuclear Physics, Research School of Physical Sciences and Engineering, The Australian National University, Canberra, ACT 2601, Australia}

\date{December 26, 2017}% - v1

\begin{abstract}
Energy dissipative processes play a key role in how quantum many-body systems dynamically evolve towards equilibrium. In closed quantum systems, such processes are attributed to the transfer of energy from collective motion to single-particle degrees of freedom; however, the quantum many-body dynamics of this evolutionary process are poorly understood. To explore energy dissipative phenomena and equilibration dynamics in one such system, an experimental investigation of deep-inelastic and fusion-fission outcomes in the $^{58}$Ni+$^{60}$Ni reaction has been carried out. Experimental outcomes have been compared to theoretical predictions using Time Dependent Hartree Fock and Time Dependent Random Phase Approximation approaches, which respectively incorporate one-body energy dissipation and fluctuations. Excellent quantitative agreement has been found between experiment and calculations, indicating that microscopic models incorporating one-body dissipation and fluctuations provide a potential tool for exploring dissipation in low-energy heavy ion collisions. 
\end{abstract}
\maketitle

The dynamic evolution of perturbed quantum many-body systems towards equilibrium is a topic of great interest in many fields, including quantum information \cite{Suter16}, condensed matter \cite{Polkovnikov11, Gring12, Aoki14, Eisert15}, and nuclear physics \cite{Hinde10, Simenel11, Aritomo12, Lacroix14, Umar15}. Energy dissipation---the transfer of energy from collective motion to internal or external degrees of freedom---shapes this dynamic evolution, playing a significant role in whether and how such complex systems achieve full equilibration. To date, a great deal of effort has focused on quantum systems in which energy dissipation is brought about via contact with an external environment (e.g., gas, photons, etc.) \cite{Barontini13, deVega17}. Much less is known about energy dissipation that arises from internal degrees of freedom \cite{Polkovnikov11, Eisert15, Clos16}.

One testing ground for the exploration of energy dissipation due to internal degrees of freedom can be found in heavy ion collisions. The nuclear collision process results in a closed composite quantum system that is isolated from external environments during the time of the collision (a timescale of several zeptoseconds, prior to particle emission), rapidly evolves towards equilibration in many degrees of freedom, and undergoes significant excitation and internal rearrangement throughout the equilibration process. Through the manipulation of collision entrance channel parameters (projectile-target combinations and energies), a range of factors with the potential to affect energy dissipation can be explored. Typical timescales for energy dissipation in such systems could in principle vary from isospin and mass equilibration times on the order of 0.3-0.5 zs \cite{Jedele17,Umar17} and $\sim$5 zs \cite{Toke85,duRietz13}, respectively.

In nuclear reactions, the observation of the total kinetic energy of the reaction products (TKE) offers a direct measure of energy dissipation. The observation of the masses of reaction products via direct or indirect methods offers a measure of system equilibration in a key degree of freedom, and can be used to explore fluctuations in reaction product masses as a function of TKE. One or both observables have often been used to explore energy dissipative outcomes in nuclear physics (see, e.g. \cite{Bromley}). In this work, we will explore the TKE and mass degrees of freedom for binary outcomes of collisions between $^{58}$Ni and $^{60}$Ni at energies near the Coulomb barrier.

We have chosen to study low-energy collisions between $^{58}$Ni and $^{60}$Ni for several reasons. First, the entrance channel is close to symmetry. This means we can focus on quantum fluctuations without taking into account macroscopic mass drift effects while avoiding the experimental difficulties that come with symmetric reactions (e.g. normalization with Mott scattering, indistinguishability of projectile-like and target-like outcomes). Second, the system is relatively heavy---meaning it truly qualifies as a composite many-body system---but is still light enough that the charge product, and thus, the Coulomb repulsion, of the system is fairly small, allowing for long contact times. Most importantly, the system is accessible via both experiment and theory, using stable heavy ion beams in the former case, and microscopic approaches employing one-body dissipation (Time-Dependent Hartree Fock, or TDHF \cite{Dirac30}) and fluctuations (Time Dependent Random Phase Approximation, or TDRPA \cite{Balian84}) in the latter. 
         
The $^{58}$Ni+$^{60}$Ni experiment was performed at the ANU Heavy Ion Accelerator Facility (HIAF) using the 14UD tandem accelerator, CUBE two-body fission spectrometer \cite{Hinde96} and two monitor detectors at 18$^\circ$ for cross section normalization.  The $^{58}$Ni beam impinged upon the 60 $\mu g$/cm$^2$-thick $^{60}$Ni target for 20 separate beam energies ranging from $\sim$194-270 MeV. The CUBE spectrometer's two large-area multiwire proportional counters (each 27.9 cm wide, 35.7 cm high) were placed at forward angles $45^\circ$ relative to the beam axis and a distance of 22.24 cm from the target. The detectors provided energy loss, time of flight, and $(x,y)$ position information with a resolution of 1 mm. From this, a full kinematic reconstruction of each two-body event was performed using the kinematic coincidence method \cite{Hinde96, Rafiei08}, providing total kinetic energy (TKE), mass ratio $M_R=M_1/(M_1+M_2)$, where $M_i$ are the masses of the two fission fragments, and scattering angle information.
\begin{figure}
\includegraphics[width=8.6cm]{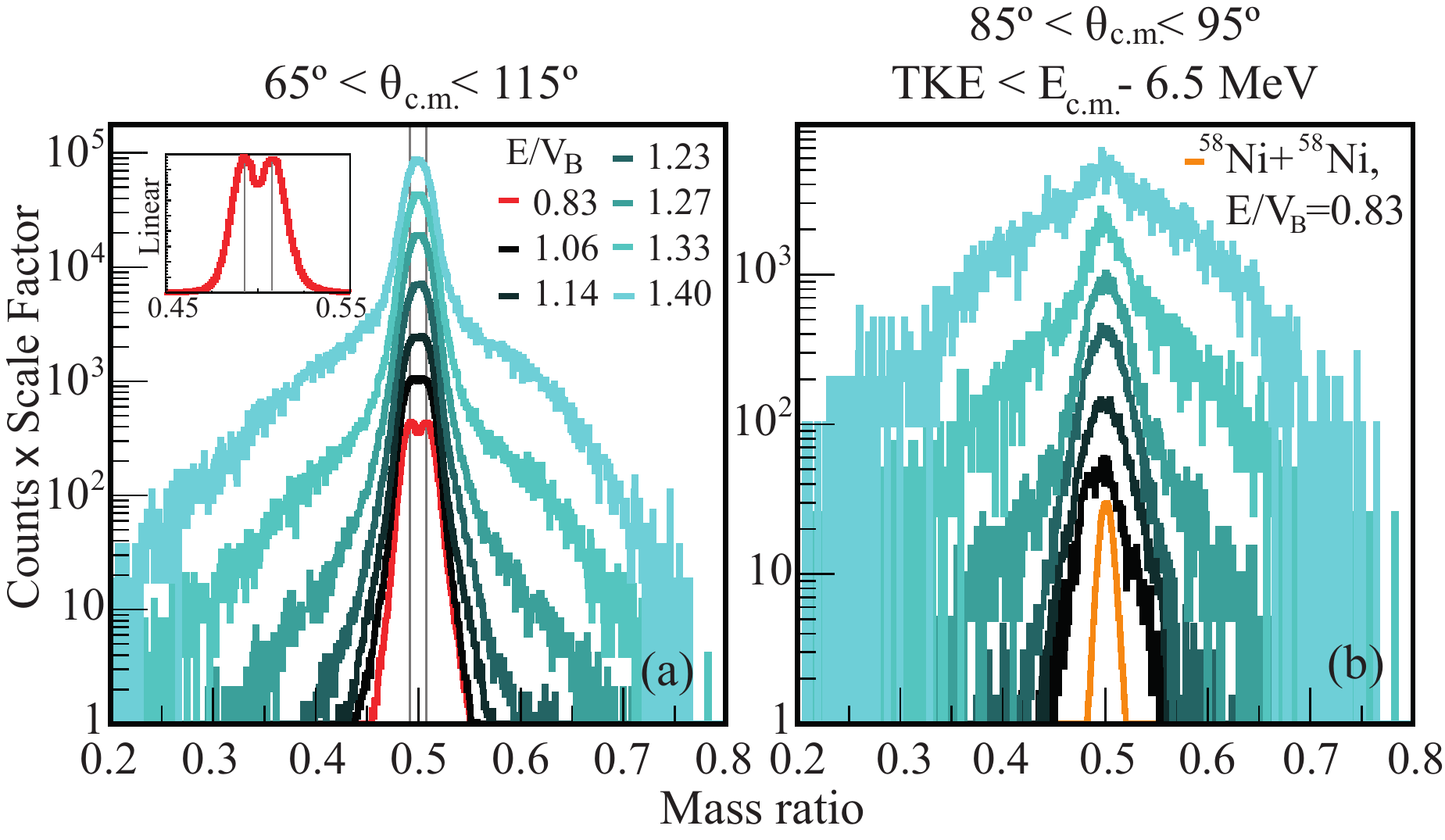}
\caption{\label{fig:mrevol} (Color online) Mass ratio distributions from the $^{58}$Ni+$^{60}$Ni reaction for a range of $E/V_B$. In (a), a $65^\circ < \theta_{c.m.} < 115^\circ$ gate has been used to ensure full detector coverage. The thin vertical lines indicate the expected mass ratios for the projectile and target nuclei. The inset shows a zoomed-in version of the $E/V_B$=0.83 mass ratio distribution on a linear scale.  In (b), a gate has been placed on TKE that excludes 99$\%$ of the Rutherford scattered events, together with a $85^\circ<\theta_{c.m.}<95^\circ$ gate that isolates mass ratios corresponding to the region where fusion-fission events are expected to be found. The orange line shows $^{58}$Ni+$^{58}$Ni data at $E/V_B=0.83$ without angle or TKE gate. No background subtraction has been applied to any of these mass distributions.} 
\end{figure}

Calibrations were performed with a $^{58}$Ni beam of 158.4 MeV bombarding (50, 60)-$\mu$g/cm$^{2}$ $^{58,60}$Ni targets. Mott or Rutherford scattering were used to calibrate the geometry of the setup, define the mass-ratio resolution of CUBE, and provide a solid angle calibration for cross section measurements. 

Figure \ref{fig:mrevol} shows the evolution of mass ratio distributions observed for $^{58}$Ni+$^{60}$Ni with $E/V_B$, or energy $E$ relative to the experimental fusion barrier $V_B$=96.87 MeV \cite{Rodriguez10} in the center-of-mass frame. In Fig. \ref{fig:mrevol} (a), an angular acceptance of $65^\circ < \theta_{c.m.} < 115^\circ$ has been used to exclude regions where detector coverage is incomplete. The lowest energy distribution (red line, $E/V_B=0.83$), exhibits a double-peaked structure as highlighted by the linear scale inset of the same data. This is consistent with expectation for elastically scattered projectile ($M_R = 0.49$) and target ($M_R = 0.51$) nuclei. As $E/V_B$ is increased, the double peak becomes a single peak due to mass equilibration, and the previously narrow mass ratio distribution develops a second, wide component.

In Fig. \ref{fig:mrevol} (b), the wide component of the mass-ratio distribution at higher energies is highlighted by applying two gates: one around $\theta_{c.m.} = 90^\circ$ that minimizes elastic and deep-inelastic events, and one below TKE = $E_{c.m.} - 6.5$ MeV that excludes $\sim$99$\%$ of the Rutherford scattering events. As beam energy increases, the narrow (wide) component of the mass ratio distribution decreases (increases) in yield. At the highest energy, almost no trace of the narrow component remains. For reference, the $^{58}$Ni + $^{58}$Ni mass ratio (orange) shows the experimental mass-ratio resolution for an individual nuclide. 
\begin{figure*}
\includegraphics[width=15cm]{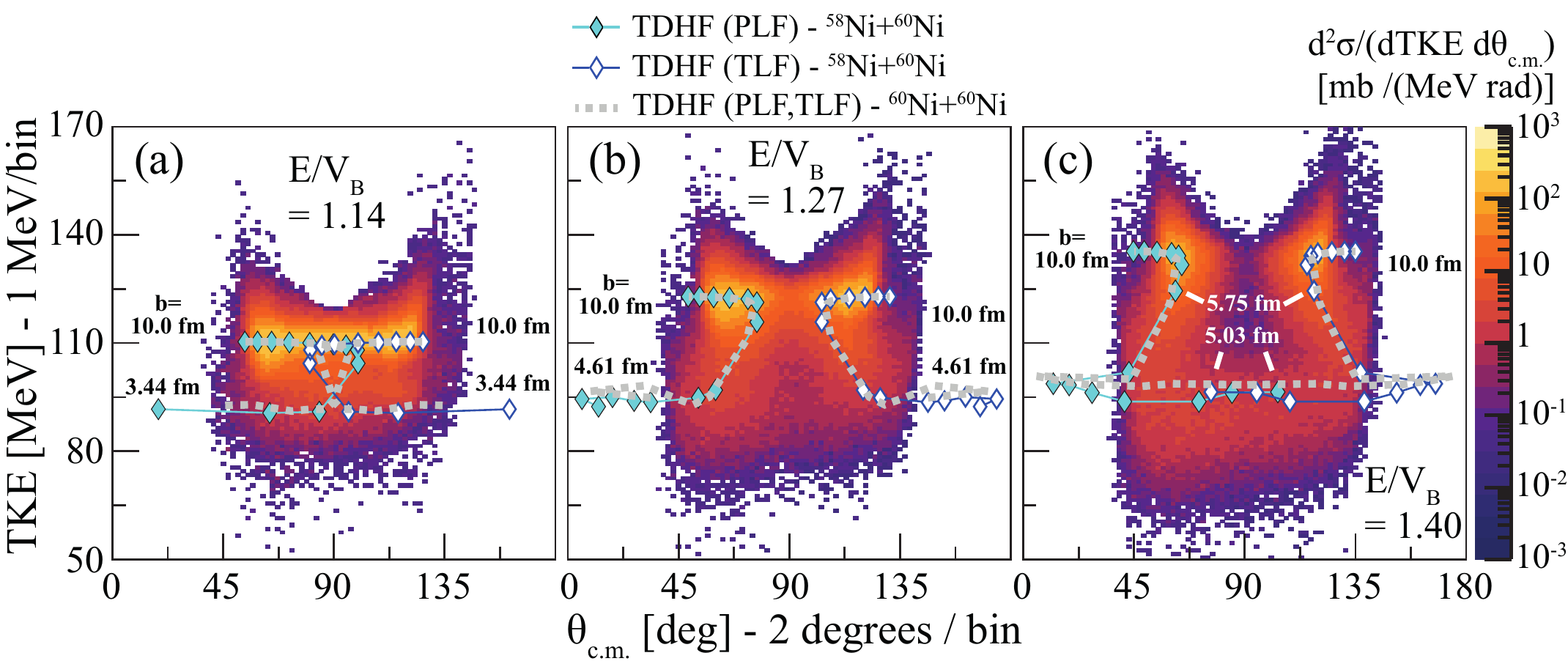}
\caption{\label{fig:thcmtke} (Color online) Scattering angle $\theta_{c.m.}$ versus total kinetic energy (TKE) differential cross section distributions $d^2\sigma/(d\theta_{c.m.} d$TKE$)$  for $^{58}$Ni + $^{60}$Ni at three different energies overlaid with ($\theta_{c.m.}$, TKE) impact parameter trajectories calculated with TDHF. (Diamond symbols) $^{58}$Ni+$^{60}$Ni, where black diamonds are projectile-like fragments (PLF), white diamonds are target-like fragments (TLF); (Dotted lines) $^{60}$Ni+$^{60}$Ni.  Differential cross sections are only a lower limit outside of the range of $65^\circ < \theta_{c.m.} < 115^\circ$ due to detector angular coverage. The impact parameter $b$ for selected points along the TDHF trajectories are shown.}
\end{figure*}	
	
To identify the various reaction outcomes contributing to the mass-ratio spectra in Fig. \ref{fig:mrevol}, the experimental TKE information must also be considered as it provides information on whether energy dissipation has occurred. In Fig. \ref{fig:thcmtke}, $\theta_{c.m.}$ versus TKE is shown for three representative energies (overlaid with calculations, to be discussed below). The highest intensity (light yellow) regions at high TKE values are consistent with elastic scattering energies. Events with lower TKE values, present in each case (for example, below $\approx $ 110 MeV in Fig. \ref{fig:mrevol} (a)), provide evidence of energy dissipative reaction processes as early as $E/V_B = 1.14$.  Such events could correspond to either deep-inelastic or fusion-fission outcomes.

The main distinguishing experimental features for deep-inelastic and fusion-fission processes here are that fusion-fission outcomes should have full energy dissipation (e.g. appear in a flat TKE band, like that seen in Fig. \ref{fig:thcmtke}(c) around $\approx$ 80 MeV) and exhibit a wide mass distribution (indicating that significant mass transfer has occurred), while deep-inelastic collisions only need to show energy dissipation relative to expected elastic scattering TKE values and may involve only a small amount of mass transfer (i.e. the incoming and outgoing reaction channel masses can be similar). From Fig. \ref{fig:mrevol}(b), it is clear that the wide mass ratio component of the data (which is consistent with fusion-fission) does not become important until $E/V_B \sim 1.27$. Based on the experimental information alone, it is unclear whether the wide mass-ratio distributions observed in Fig. \ref{fig:mrevol} are solely due to fusion-fission, a deep-inelastic process, or some combination of the two. 

To explore the relationship between experimental observables ($\theta_{c.m.}$, TKE) and impact parameter $b$, and to examine the origin of the wide mass ratio component, microscopic approaches TDHF and TDRPA have been used. TDHF is a mean-field approach incorporating one-body dissipation that has  previously been used to explore dynamical reaction processes such as quasifission \cite{Simenel12, Simenel12a, Wakhle14, Oberacker14, Umar15, Hammerton15, Sekizawa16, Umar16}.  In this work, a 3D TDHF code \cite{Sekizawa13, Sekizawa13e}, employing the SLy4d parametrization \cite{Kim97} of the Skyrme energy-density functional \cite{Skyrme56} has been used to explore reaction outcomes for $^{58,60}$Ni+$^{60}$Ni at the three representative energies shown in Fig. \ref{fig:thcmtke}.  The code has been extended and also used for the TDRPA calculations.

TDHF calculations were performed for a reaction timescale of 13 zeptoseconds, and the minimum impact parameter for each calculation set was chosen to correspond to the minimum value at which the di-nuclear system was observed to reseparate into two components over this timescale. Below these minimum impact parameters, it is assumed that fusion is the most probable outcome---an assumption that is supported by calculated moments of inertia for impact parameters below the minimum range, which indicated convergence towards a compact system (see Supplemental Material \cite{SM}). The maximum impact parameter for all calculations was chosen to be 10 fm.  In Fig. \ref{fig:thcmtke}, each experimental dataset has been overlaid with the results of these TDHF calculations. Each diamond pair represents the calculated average $\theta_{c.m.}$ and TKE value obtained from a given impact parameter calculation for the $^{58,60}$Ni+$^{60}$Ni reactions. Impact parameters corresponding to selected points along these trajectories in ($\theta_{c.m.}$, TKE)-space have been noted; we will henceforth call these trajectories in ($\theta_{c.m.}$, TKE)-space ``impact parameter trajectories''. Both reactions yield a similar evolution in ($\theta_{c.m.}$,TKE)-space. For the three representative calculations, the largest impact parameters (at high TKE values) result in little energy dissipation (yielding points near elastic scattering outcomes), while the smallest impact parameters (at low TKE values) exhibit significant energy dissipation.  The impact parameters for each calculation are not evenly spaced to capture the rapid change in dynamics in the small impact parameter range. 

As one can see from Fig. \ref{fig:thcmtke} (a-c), the TDHF impact parameter trajectories follow the trends in the data quite well over the full experimental angular range.  Outcomes for small impact parameters exhibit increased energy dissipation and di-nuclear system rotation before re-separation relative to outcomes for large impact parameters. In the calculations, we define contact time as the time during which the density overlap between the fragments is $\rho>$0.001 fm$^{-3}$ (selected to provide a smooth evolution in contact time as a function of $b$, as discussed in \cite{SM}). This generally increases as $b$ decreases, thus allowing more time for both rotation and energy dissipation to occur. 

\begin{figure}
\includegraphics[width=8.6cm]{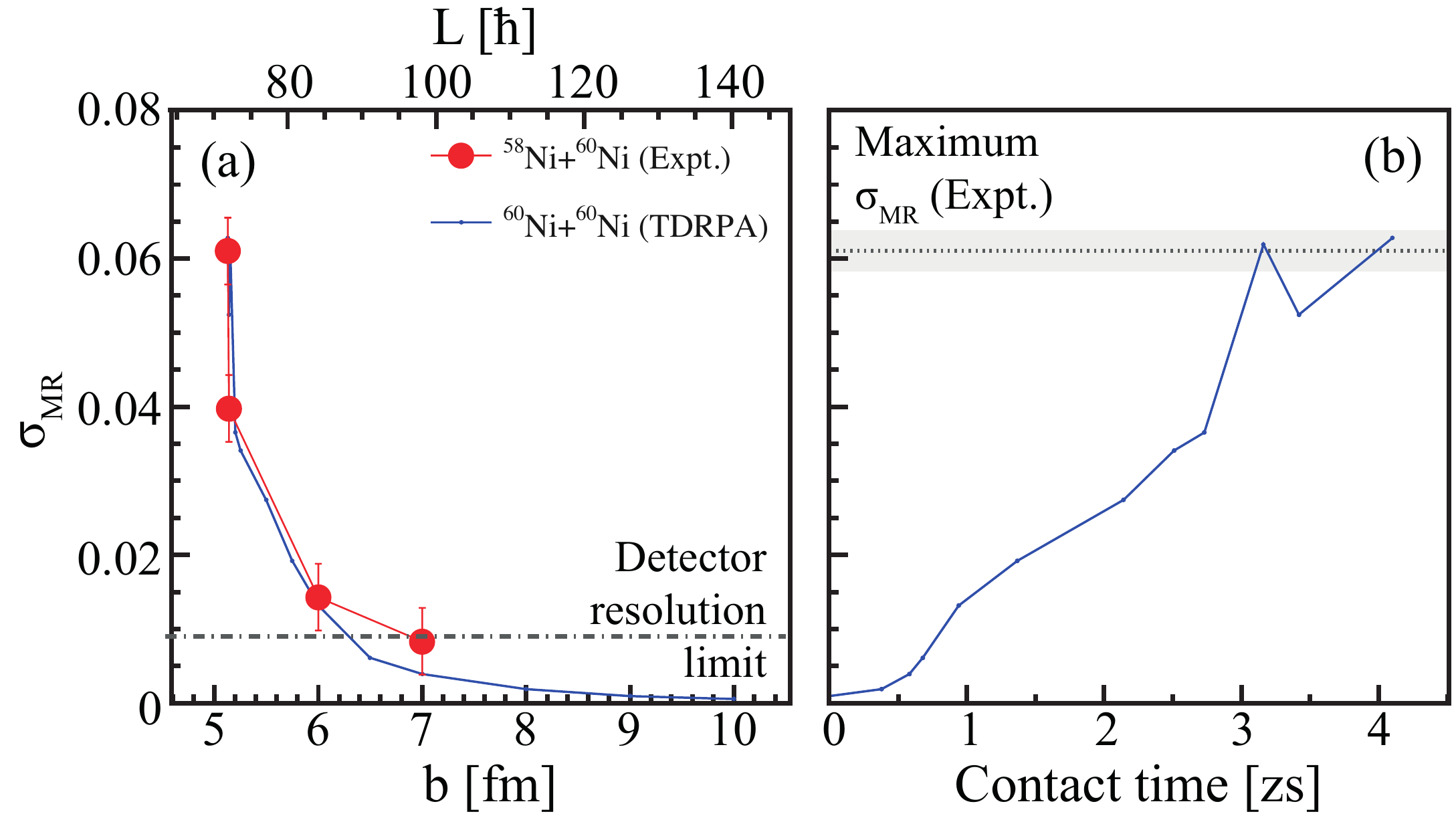}
\caption{\label{fig:bcalc} (Color online) (a) TDRPA and observed $\sigma_{MR}$ values are plotted as a function of calculated impact parameter $b$ for $E/V_B=1.40$. For the experimental data, $\sigma_{MR}$ is obtained from a Gaussian fit of the mass ratio within a ($\theta_{c.m.}$,TKE) gate corresponding to the equivalent trajectory given by the symmetric calculation; the impact parameter for each point is assigned based on the symmetric trajectory. The error bars on the experimental points and the gray dashed line show the detector mass resolution obtained from the $\sigma_{MR}$ of the below-barrier $^{58}$Ni+$^{58}$Ni data, corresponding to the detector mass-ratio resolution for a single atomic mass. Because data points have been selected along the projectile-like branch of the trajectory, points at high TKE values (large $b$) are expected to approach the resolution limit. (b) TDHF contact time versus $\sigma_{MR}$ from TDRPA. The dotted line indicates maximum $\sigma_{MR}$ from experiment with the shaded region representing the detector mass-ratio resolution.} 
\end{figure}

So far, we have found that the reaction outcomes are characterized by (i) strong TKE-angle correlations, and (ii) a transition between narrow and wide components in the fragment mass distributions. The next step is to investigate the correlation between energy dissipation and mass fluctuations in the fragments. Experimentally, we can achieve this by examining the evolution of $\sigma_{MR}$, the standard deviation of a Gaussian fit to the mass ratio distribution along the TDHF impact parameter trajectories. To compare this to theory, we need to first examine the ability of microscopic theories to reproduce this evolution of $\sigma_{MR}$ along the impact parameter trajectories shown in Fig. \ref{fig:thcmtke}.

Because of its mean-field nature, TDHF is optimized to calculate the average of one-body observables (e.g., fragment mass, charge), but underestimates their fluctuations \cite{Dasso79}. A realistic estimate of the latter can be obtained with TDRPA \cite{Balian84}, an extension of TDHF that successfully reproduced mass distribution widths in $^{40}$Ca+$^{40}$Ca \cite{Simenel11}. The TDRPA dispersion formula can be derived from the stochastic mean-field approach \cite{Ayik08}. TDRPA is used to calculate fluctuations in particle number $A$ in the fragments, which can be interpreted as a standard deviation $\sigma_A = (\overline{A^2}-\overline{A}^2)^{1/2}$ only in the case of symmetric collisions. $\sigma_A$ can then be used to compute the calculated equivalent of $\sigma_{MR}$ by dividing by the compound nucleus mass number. As Fig. \ref{fig:thcmtke} showed, the TDHF trajectories for $^{58,60}$Ni+$^{60}$Ni were found to be very similar; therefore, the calculated $^{60}$Ni+$^{60}$Ni trajectories and $\sigma_{MR}$ values will be used in the discussion below. For completeness, we have also computed fragment mass fluctuations in $^{58}$Ni+$^{60}$Ni with TDRPA, to see if the asymmetry is small enough to allow an interpretation of the TDRPA mass fluctuations as $\sigma_A$ in this system. The details of this calculation can be found in \cite{SM}.

To compare calculations to observation, a series of scattering angle and TKE gates were placed on the data along the TDHF trajectories identified in Fig. \ref{fig:thcmtke}. Each gate was centred on the $\theta_{c.m.}$ and TKE value resulting from each TDHF impact parameter calculation where detector coverage was complete. The gate widths were chosen as follows: (i) TKE gate widths were $\approx$5 MeV, corresponding to the observed full-width-half-max of the TKE for elastic scattering events in the $^{58}$Ni+$^{60}$Ni calibration run at $E/V_B=0.83$, and (ii) $\theta_{c.m.}$ gate widths were $5^\circ$, as this width yielded reasonable statistics in the mass distributions. The resulting experimental $\sigma_{MR}$ values are shown as a function of TDHF impact parameter $b$ in Fig. \ref{fig:bcalc} (a) for the highest energy measurement. Here, $\sigma_{MR}$ decreases sharply over a narrow impact parameter range. The largest $\sigma_{MR}$ values, found at the smallest impact parameters, correspond to mass ratio widths consistent with those observed for fusion-fission outcomes in heavier systems (see, e.g., \cite{Knyazheva07, Rafiei08, Williams13}).

The TDRPA results for $^{60}$Ni+$^{60}$Ni are also shown in Fig. \ref{fig:bcalc} (a). The agreement between the theoretical (mass-symmetric) and experimental $\sigma_{MR}$ is remarkable, considering the fact that the only input of TDRPA (and TDHF) is the choice of the effective interaction. This conclusion holds for all three energies. For the reaction at $E/V_B$=1.14 and minimum impact parameter $b$ = 3.8 fm, $\sigma_{MR, expt}= 0.027(9)$ and $\sigma_{MR, calc}= 0.017$; for the reaction at $E/V_B$=1.27 and minimum impact parameter $b$ = 5.0 fm, $\sigma_{MR, expt}= 0.025(9)$ and $\sigma_{MR, calc}= 0.017$. Experimental errors on $\sigma_{MR}$ given here originate from the detector resolution, as determined from the $^{58}$Ni+$^{58}$Ni calibration data. In both cases, experimental $\sigma_{MR}$ values then decrease quickly to the CUBE resolution limit as $b$ is increased.

The TDHF and TDRPA calculations also reveal something about the wide mass component of the highest energy reaction that the experimental data cannot: the reaction timescale. As shown in Fig. \ref{fig:bcalc} (b), the widest calculated $\sigma_{MR}$ values result from reactions with longer contact times ($\gtrsim 3$ zeptoseconds). These calculated mass distribution widths are consistent with those observed for the wide mass component of our experimental data, supporting the idea that this wide mass component results from systems that have more time to undergo mass exchange prior to reseparation. This suggests that the wide mass ratio component of the experimental data could originate from both deep-inelastic and fusion-fission processes, where the $\sim$4 zeptosecond timescale applies to the former process and a much longer ($> 10^{-20}$ s) timescale to the latter. Remarkably, the fact that the largest calculated mass ratio widths could be consistent with those expected for fusion-fission, where full mass equilibration has by definition occurred, also suggests that full mass equilibration can be achieved over a timescale of only $\sim$4 zeptoseconds for this reaction. This timescale range is consistent with the 5-10 zs timescales observed for quasifission outcomes in some heavier systems \cite{Toke85,duRietz11,duRietz13}.

In summary, this work illustrates the strong quantitative agreement between experiment and a quantum many-body approach including only a one-body dissipation mechanism for low-energy heavy ion collisions, supporting the idea that such approaches are highly appropriate for exploring energy dissipative processes in composite quantum many-body systems. The 4-zeptosecond timescale for the mass equilibrated deep-inelastic outcome, as calculated by TDRPA, suggests that the microscopic mechanisms driving mass equilibration can operate very rapidly in low-energy heavy ion reactions.

\textbf{Acknowledgements:} The authors thank the technical staff of the ANU Heavy Ion Accelerator Facility for their essential support during the experiments.  The authors also acknowledge support from the Australian Research Council through Grants DE140100784, FT120100760, FL110100098, DP130101569, DP140101337, and DP160101254, from the NCRIS program for accelerator operations, and from the Polish National Science Centre (NCN) Grant, Decision No. DEC-2013/08/A/ST3/00708. This work used computational resources of the HPCI system (HITACHI SR16000/M1) provided by Information Initiative Center (IIC), Hokkaido University, through the HPCI System Research Projects (Project IDs: hp140010, hp150081, hp160062, and hp170007).

% ============================================================================
% ================================ SUPPLEMENT ==================================
% ============================================================================

\vspace{5mm}
\begin{center}
{\bf Supplemental Material for:}
{\bf ``Exploring Zeptosecond Quantum Equilibration Dynamics: From Deep-Inelastic to Fusion-Fission Outcomes in $^{58}$Ni+$^{60}$Ni Reactions''}\\
\end{center}
\begin{small}
\noindent
In this Supplemental Material, computational details of TDHF and TDRPA calculations are described.
Supplemental results that show the reaction dynamics, the time-evolution of moment of inertia in
fusing systems, and contact times are presented. We point out the difficulty of using TDRPA calculations
for asymmetric systems. A short discussion on fluctuations and correlations in deep-inelastic reactions
is provided. 
\end{small}

\section{Computational details}

For the TDHF calculations, we used a computational code developed in University
of Tsukuba \cite{KS_KY_MNT}. In the code, single-particle orbitals are represented
in three-dimensional Cartesian coordinates without any symmetry restrictions and
with hard boundary conditions. The coordinate space is discretized into uniform grids
with a mesh spacing of 0.8~fm. Spatial derivatives are evaluated by an 11-point
finite-difference formula. For time evolution, a 4th-order Taylor-expansion method
is used with $\Delta t=6.7\times10^{-4}$~zs (1~zs~$=10^{-21}$~sec).
The Coulomb potential is computed employing Fourier transformations.

For the energy density functional, we used the Skyrme SLy4d parameter set \cite{SLy4d}.
TDHF calculations were performed for the $^{58,60}$Ni+$^{60}$Ni reactions at $E/V_B
=1.14$, 1.27, and 1.40. The Hartree-Fock ground state of $^{58}$Ni is slightly deformed
in a prolate shape with $\beta \approx 0.12$, while that of $^{60}$Ni is of spherical shape.
For the $^{58}$Ni+$^{60}$Ni reaction, the symmetry axis of $^{58}$Ni was always set
perpendicular to the reaction plane. The initial separation distance between projectile and
target nuclei for TDHF calculations was set to 24~fm along the collision axis. When binary reaction
products were generated, time evolution was continued until $t=t_{f}$ after the collision,
at which point the relative distance between the reaction products reaches 26~fm. When
two nuclei merged, forming a composite system, time evolution was continued up to around
13~zs. If the composite system kept a compact shape (as quantified by moments of inertia,
see Sec.~\ref{Sec:MoI}) within this period, we regarded it as a fusion reaction. The minimum
impact parameter $b_{\rm min}$ inside which fusion reactions take place was obtained by
repeating TDHF calculations with an 0.01-fm impact parameter step. For the $^{58}$Ni+$^{60}$Ni
reaction at $E/V_{\rm B}=1.14$, 1.27, and 1.40, respectively, we found $b_{\rm min}=3.44$, 4.61,
and 5.03~fm, while those for the $^{60}$Ni+$^{60}$Ni reaction were found to be $b_{\rm min}
=3.74$, 4.79, and 5.13~fm, respectively.

The TDRPA calculations were carried out by extending the computational code that was
used for the TDHF calculations. Here we recall only an essential part of the computations,
and we refer Refs.~\cite{TDRPA,Simenel(review)} for details of the technique. The fluctuation
and correlation can be computed based on TDRPA with minor modifications of an existing TDHF
code. Generally, the correlation between two one-body observables is given by
\begin{equation}
\sigma_{XY} = \sqrt{\bigl<\hat{X}\hat{Y}\bigr> - \bigl<\hat{X}\bigr>\bigl<\hat{Y}\bigr>},
\end{equation}
where $\hat{X}$ and $\hat{Y}$ denote arbitrary one-body operators. The fluctuation of
a one-body observable corresponds to the $\hat{X}=\hat{Y}$ case. To compute fluctuations
and correlations of neutron, charge, and mass numbers of a reaction product, the operators
are taken as the number operator for a spatial region $V$ in which the reaction product exists.
At the time $t=t_{f}$ after collision, at which one wants to evaluate those fluctuations and
correlations, one can transform single-particle orbitals as
\begin{equation}
\varphi_\alpha^{(X)}(\br\sigma q,t_{f}; \varepsilon)
= \exp[-\zI\varepsilon\,\xi_X \Theta_V(\hat{\bs{r}})]\,\psi_\alpha(\br\sigma q,t_{f}),
\label{Eq:varphi}
\end{equation}
where $\varepsilon$ is an infinitesimal constant. The step function $\Theta(\bs{r})$ is equal to
1 within the subspace $V$, and 0 elsewhere. $\psi_\alpha(\bs{r}\sigma q,t_f)$ denotes the
$\alpha$th single-particle orbital at time $t=t_f$ with spatial, spin, and isospin coordinates $\bs{r}$,
$\sigma$, and $q$, respectively. Note that $\xi_X=1$ for $\hat{X}=\hat{N}_V=\hat{N}_V^{(n)}
+\hat{N}_V^{(p)}$, while $\xi_X=\delta_{qq'}$ for $\hat{X}=\hat{N}_V^{(q')}$ ($q'=n$ or $p$).
For the transformed single-particle orbitals, Eq.~(\ref{Eq:varphi}), a ``backward" TDHF evolution
from $t=t_{f}$ to $t_{i}$ is performed, where $t_{i}$ is the initial time of the TDHF calculation.
From the backward evolution, we obtain $\varphi_i^{(X)} (\br\sigma q,t_{f}; \varepsilon)
\rightarrow \varphi_i^{(X)}(\br\sigma q,t_{i}; \varepsilon)$. The fluctuations and correlations
can then be evaluated as
\begin{equation}
\sigma_{XY}(t_{f})
= \sqrt{\frac{\eta_{00}(t_{i}) + \eta_{XY}(t_{i}) - \eta_{0X}(t_{i}) - \eta_{0Y}(t_{i})}{2\varepsilon^2}},
\label{Eq:sigma_XY}
\end{equation}
where
\begin{eqnarray}
&&\eta_{XX'}(t_{i}) \0\\[0.5mm]
&&\,= \sum_{\alpha\beta}\,\Bigg| \sum_\sigma \int
\varphi_\alpha^{(X)}(\bs{r}\sigma q,t_i;\varepsilon)
\varphi_\beta^{(X')}(\bs{r}\sigma q,t_i;\varepsilon)
\,d\bs{r} \,\Bigg|^2.
\end{eqnarray}
The subscript `$0$' in Eq.~(\ref{Eq:sigma_XY}) means that $\varphi_\alpha^{(0)}(t_{i})$ is computed
by the backward evolution using $\hat{X}=0$, namely, without multiplying the phase factor at
$t=t_{f}$. The usage of $\eta_{00}$ improves numerical accuracy. In practice, the spatial region $V$
is taken as a sphere with a radius of 13\,fm around the center-of-mass of the projectile-like fragment.
The infinitesimal constant $\varepsilon$ has to be small enough to ensure a convergence of the calculation.
The convergence was confirmed taking $\varepsilon=10^{-2},10^{-3}$, and $10^{-4}$. We used
$\varepsilon=10^{-4}$ for the $E/V_B=1.27$ and 1.40 cases, while $\varepsilon=10^{-3}$ was
used for the $E/V_B=1.14$ case.

\section{Reaction dynamics}

\begin{figure}[t]
   \begin{center}
   \includegraphics[width=\columnwidth]{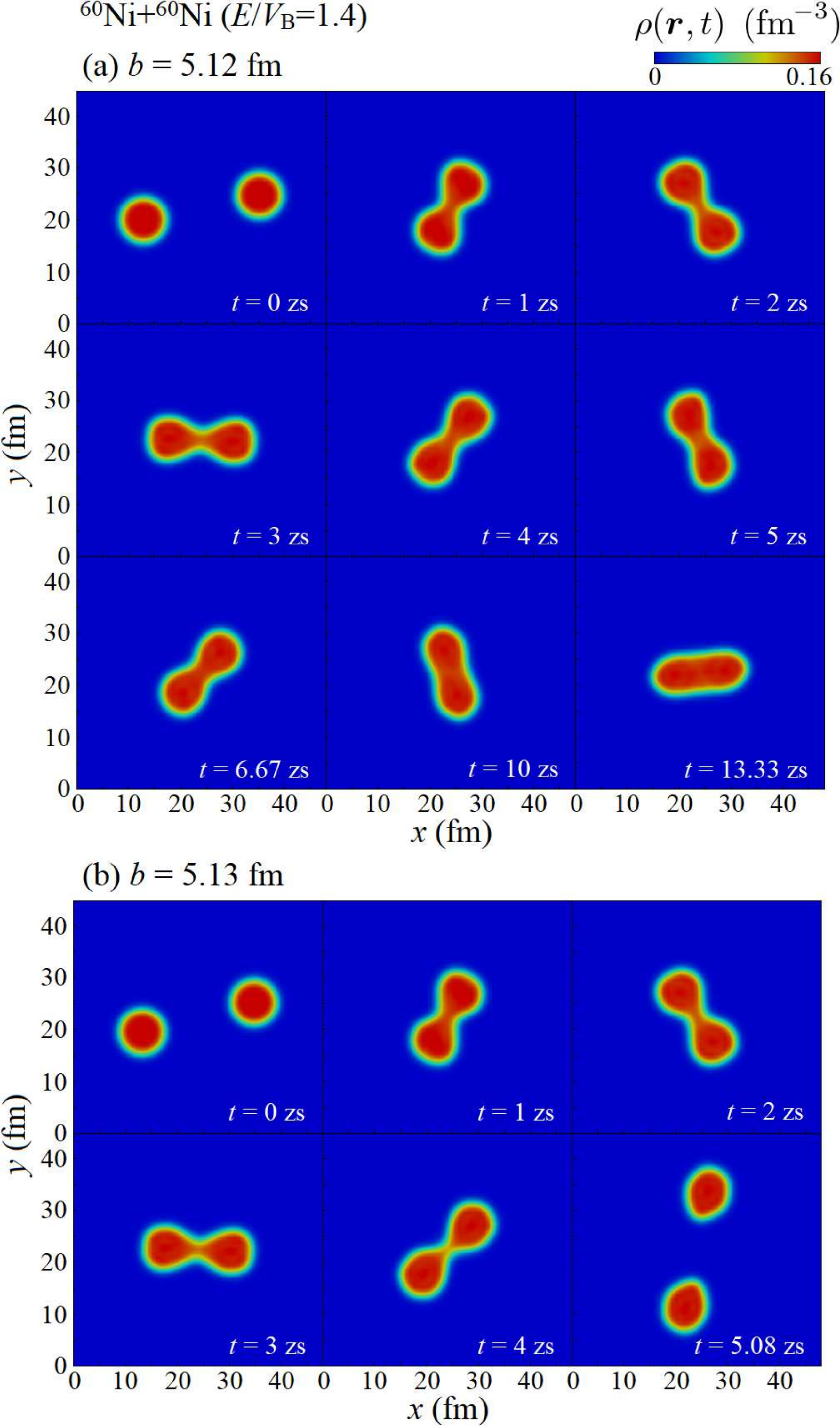}
   \end{center}\vspace{-4mm}
   \caption{
   The density of colliding nuclei at various times obtained from the TDHF
   calculations for the $^{60}$Ni+$^{60}$Ni reaction at $E/V_B=1.40$.
   Upper panels (a) show the $b=5.12$~fm case, while lower panels (b)
   show the $b=5.13$~fm case.
   }
   \label{FIG:rho(t)}
\end{figure}

Here we present typical reaction dynamics observed in the TDHF calculations.
Figure~\ref{FIG:rho(t)} shows the contour plots of the density of colliding nuclei in
the reaction plane at various times for the $^{60}$Ni+$^{60}$Ni reaction at $E/V_B=1.40$.
In upper panels (a), dynamics for the $b=5.12$\,fm case is shown, which we regarded
as a fusion reaction (see discussion in Sec.~\ref{Sec:MoI}). Note that the times are
not uniformly changed in the bottom panels in (a). In lower panels (b), dynamics for
the $b=b_{\rm min}=5.13$ fm case is shown, which corresponds to a binary reaction.
Up to $t=3$~zs, the dynamics is very similar in both two cases. In the $b=5.12$\,fm case,
the system still holds clear dinuclear structure at $t=4$~zs, and then it does not reseparate
for a long time, showing a mononuclear shape without clear neck structure ($t=13.33$~zs).
On the other hand, for the $b=5.13$\,fm case, the system evolves toward reseparation,
as can be seen from the thinner neck structure at $t=4$~zs. The right-bottom frame corresponds
to $t=t_f$ after the collision, at which point various quantities are computed.

%&&&&&&&&&&&&&&&&&&&&&&&&&&&&&&&&&&&&&&&&&&&&&&&&&&
%&&&&&&&&&&&&&&&&&&&&&&&&&&&&&&&&&&&&&&&&&&&&&&&&&&

An important quantity that characterizes the reaction dynamics is the total kinetic
energy (TKE) of the fragments, which can be a direct measure of the energy
dissipation. At $t=t_f$ after the collision [\textit{e.g.} the right-bottom panel
of Fig.~\ref{FIG:rho(t)}\,(b)], we evaluate the total kinetic energy as
\begin{equation}
{\rm TKE} = \frac{1}{2}\mu\bs{\dot{R}}^2 + \frac{Z_1Z_2e^2}{|\bs{R}|},
\end{equation}
where $\mu = m A_1A_2/(A_1+A_2)$ is the reduced mass ($m$ is the mass
of a nucleon), $A_i$ and $Z_i$ ($i=1,2$) are the mass and charge of each fragment,
$\bs{R}$ and $\bs{\dot{R}}$ are a relative vector connecting center-of-mass
positions of the fragments and its time derivative, respectively. From the information
of the position and the velocity of the fragments, the scattering angle (an asymptotic
value for $R\rightarrow\infty$ including the correction of the Coulomb potential for
$t>t_f$) can also be evaluated. Those quantities will be given in Sec.~\ref{Sec:tables}.

\section{Moment of inertia}\label{Sec:MoI}

TDHF-TDRPA simulations only give access to the early stage of the formation of the compound nucleus. These calculations are generally run over few 10 zs (13 zs in this work). Beyond these relatively short timescales, neglecting the collision terms of the residual interaction would not be valid. As a result, TDHF calculations, as well as their extensions, cannot be run long enough to allow for investigation of the competition between evaporation and fission of the compound nucleus. Instead, we determine whether an outcome (on average) leads to fusion by looking at how the moment of inertia of the dinuclear system evolves as a function of time.

In order to examine stability of a fusing system, we investigate its effective moment of inertia,
$\mathcal{I}_{\rm eff}$. The effective moment of inertia is given by \cite{MOI}
\begin{equation}
\frac{1}{\mathcal{I}_{\rm eff}} = \frac{1}{\mathcal{I}_{\parallel}} - \frac{1}{\mathcal{I}_{\perp}},
\end{equation}
where $\mathcal{I}_{\parallel(\perp)}$ denotes an eigenvalue of the moment-of-inertia tensor,
\begin{equation}
\mathcal{I}_{ij} = m\int \rho(\br,t)\,[r^2\delta_{ij} - x_i x_j]\,d\br,
\end{equation}
where $x_{i=1,2,3}$ denote $x$, $y$, and $z$ coordinates, respectively. $\mathcal{I}_{\parallel}$
corresponds to the smallest eigenvalue, while $\mathcal{I}_{\perp}$ corresponds to average
of the other two eigenvalues.

%&&&&&&&&&&&&&&&&&&&&&&&&&&&&&&&&&&&&&&&&&&&&&&&&&&
\begin{figure}[t]
   \begin{center}
   \includegraphics[width=8.6cm]{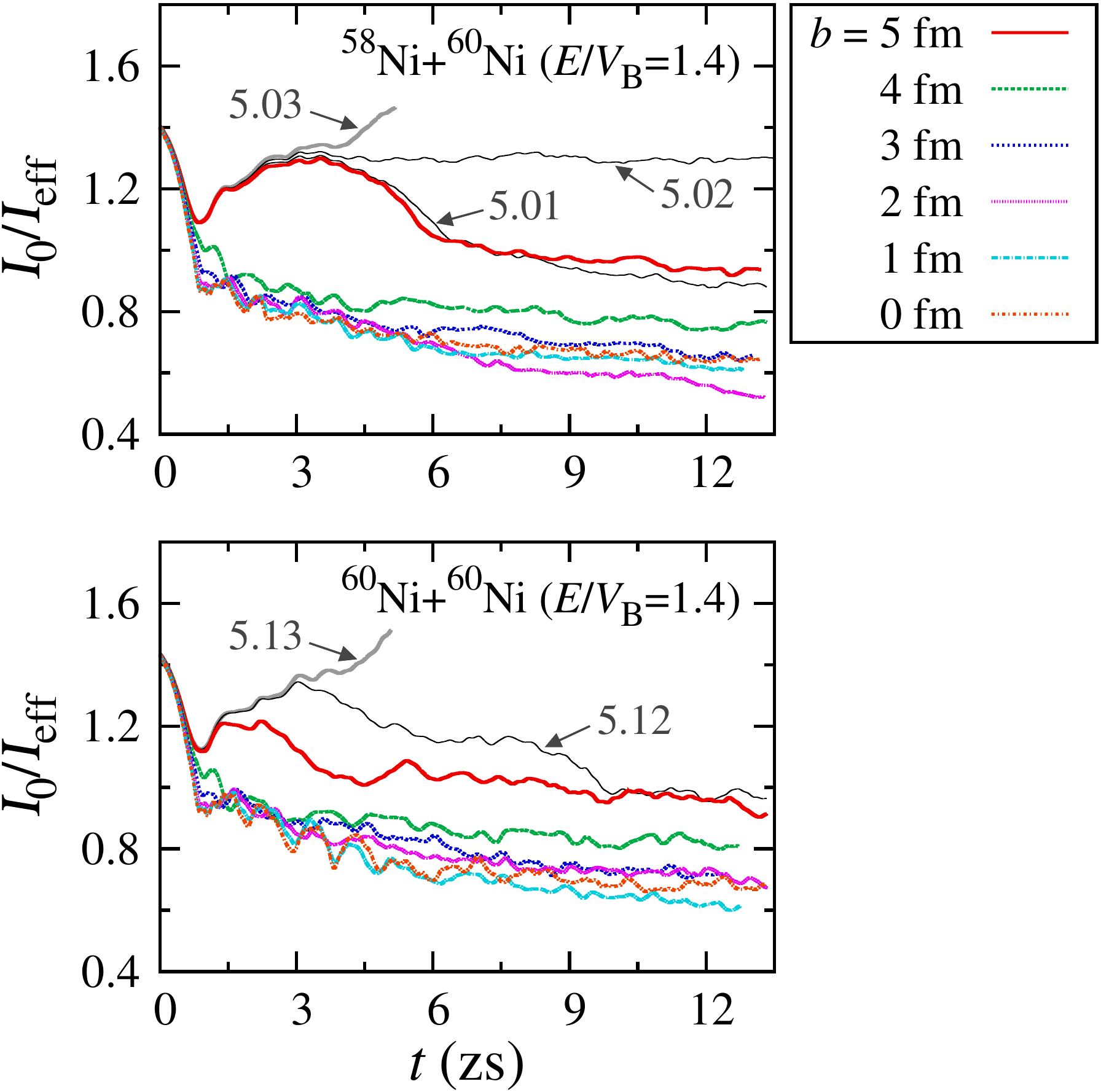}
   \end{center}\vspace{-4mm}
   \caption{
   Time evolution of the effective moment of inertia, $\mathcal{I}_0/\mathcal{I}_{\rm eff}(t)$,
   for the $^{58}$Ni+$^{60}$Ni (top) and $^{60}$Ni+$^{60}$Ni (bottom)
   reactions ($E/V_B=1.40$) for various impact parameters.
   }
   \label{FIG:MoI}
\end{figure}
%&&&&&&&&&&&&&&&&&&&&&&&&&&&&&&&&&&&&&&&&&&&&&&&&&&

In Fig.~\ref{FIG:MoI}, we show the time evolution of a ratio, $\mathcal{I}_0/\mathcal{I}_{\rm eff}(t)$,
in the $^{58}$Ni+$^{60}$Ni (top) and $^{60}$Ni+$^{60}$Ni (bottom) reactions for the highest
energy case ($E/V_{\rm B}=1.40$) at various impact parameters. $\mathcal{I}_0$ corresponds to
the moment of inertia of a spherical nucleus, $\mathcal{I}_0 = \frac{2}{5}mAR_0^2$, where $A$
is the mass number of the composite system and $R_0=1.225\,A^{1/3}$.

For $^{58}$Ni+$^{60}$Ni ($^{60}$Ni+$^{60}$Ni), we observed binary reaction products
for $b\ge5.03$~fm ($b\ge5.13$~fm). As shown in the figure, for $b \le 4$~fm cases, the composite
system becomes a compact shape within relatively short time, $t\lesssim1$\,zs. Then, the system
persists in a compact shape, as one can see from the convergent behavior of the moment of inertia in
Fig.~\ref{FIG:MoI}. At impact parameters very close to the border between capture and binary
reactions, the moment of inertia first increases up to around $t=3\mbox{--}4$~zs, then decreases,
showing a convergent trend [see also Fig.~\ref{FIG:rho(t)}\,(a)] (strictly speaking, except
$b=5.02$~fm case in $^{58}$Ni+$^{60}$Ni, where the moment of inertia is not converging,
but remains almost constant). From the convergent behavior of the moment of inertia,
we consider that the average outcome of those reactions would correspond to fusion.

\section{Contact time}

%&&&&&&&&&&&&&&&&&&&&&&&&&&&&&&&&&&&&&&&&&&&&&&&&&&
\begin{figure}[t]
   \begin{center}
   \includegraphics[width=7.0cm]{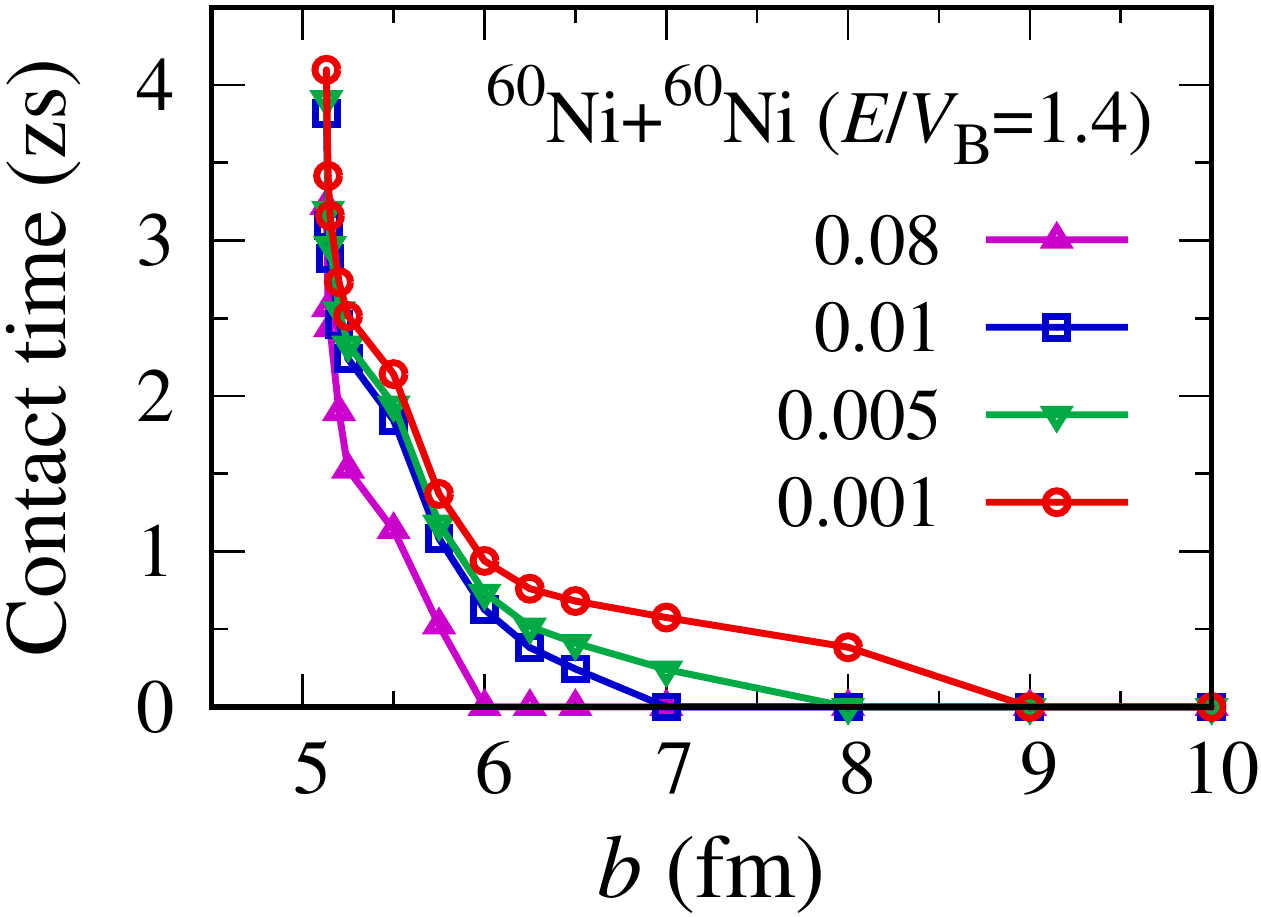}
   \end{center}\vspace{-4mm}
   \caption{
   Contact times with different criteria ($\rho_c>0.08$, 0.01, 0.005, 0.001~fm$^{-3}$)
   for the $^{60}$Ni+$^{60}$Ni reaction ($E/V_B=1.40$) are shown
   as a function of the impact parameter.
   }
   \label{FIG:t_contact}
\end{figure}
%&&&&&&&&&&&&&&&&&&&&&&&&&&&&&&&&&&&&&&&&&&&&&&&&&&

The TDHF approach allows the evaluation of contact (sticking or interaction) time of two nuclei
directly from the time evolution of the colliding system. From a TDHF calculation, one obtains the total density
of the system at each time, $\rho(\bs{r},t)$. By monitoring the density between two colliding
nuclei, we can measure how long they stick together, forming a dinuclear system connected
with a neck structure.

In Fig.~\ref{FIG:t_contact}, we show the contact time in the $^{60}$Ni+$^{60}$Ni reaction
at $E/V_{\rm B}=1.40$ as a function of the impact parameter. The contact time is shown in
zeptoseconds, and is defined as a time duration in which the lowest total density between the
colliding nuclei exceeds a critical value, $\rho>\rho_c$. In the figure, contact times obtained with
four different critical densities, $\rho_c=0.001$, 0.005, 0.01, and 0.08~fm$^{-3}$ are presented.

From the figure, one can see that the contact time is zero for trajectories at large impact
parameters, $b\gtrsim9$~fm. As the impact parameter decreases, the contact time increases
sharply, reaching the maximum value ($\approx4$~zs), at the border between fusion and binary
reactions. We find that a smaller value of the critical density provides a contact time which is more
sensitive to a formation of a dilute neck ($b\approx6$--7~fm). As we observe a smooth decrease
of the contact time as a function of the impact parameter for the $\rho_c=0.001$~fm$^{-3}$ case,
we employed this criteria in the present work.

\section{Remarks on TDRPA results}

%&&&&&&&&&&&&&&&&&&&&&&&&&&&&&&&&&&&&&&&&&&&&&&&&&&
\begin{figure}[t]
   \begin{center}
   \includegraphics[width=7.5cm]{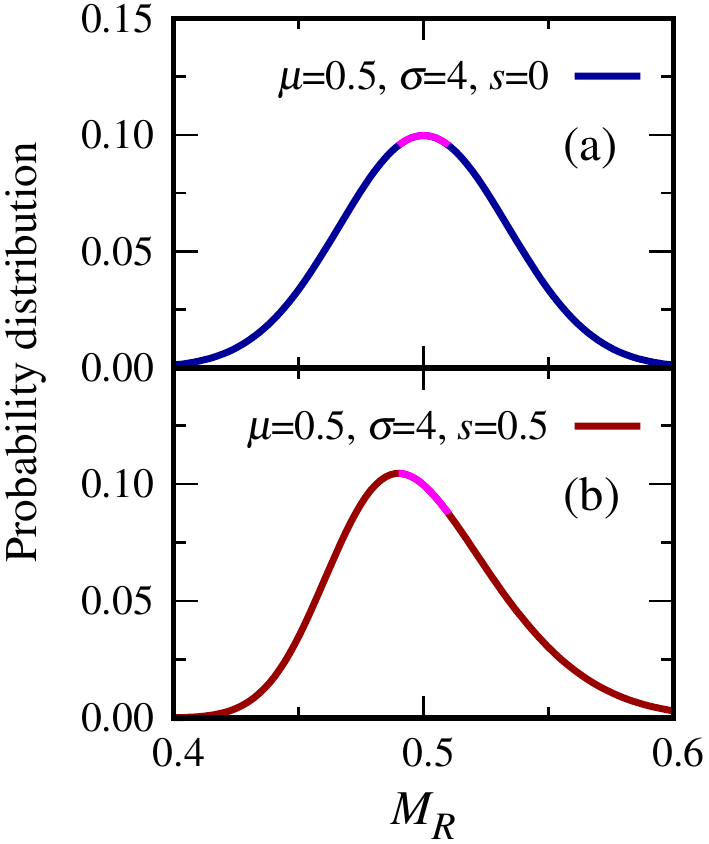}
   \end{center}\vspace{-4mm}
   \caption{
   Graphical interpretation of the failure of TDRPA calculations for the quasi-symmetric
   $^{58}$Ni+$^{60}$Ni reaction. (a) A Gaussian distribution around $\mu=0.5$ with
   standard deviation $\sigma=4$. (b) A skew-normal distribution with mean $\mu=0.5$,
   standard deviation $\sigma=4$ and skewness $s=0.5$. Curvature of the distribution
   around the mean value (highlighted by pink) is detected in TDRPA calculations (see text
   for details). Horizontal axis is rescaled like the mass width to have correspondence with
   our analyses.
   }
   \label{FIG:skewness}
\end{figure}
%&&&&&&&&&&&&&&&&&&&&&&&&&&&&&&&&&&&&&&&&&&&&&&&&&&

In this Letter, we have examined dissipation and equilibration mechanisms in the
$^{58}$Ni+$^{60}$Ni reaction. If one applies TDRPA to the $^{58}$Ni+$^{60}$Ni
reaction, however, fluctuations and correlations can be unphysically large due to a fundamental
problem that lies in the present formulation of TDRPA. The fluctuations and correlations
of one-body operators, $\hat{Q}_i$, can be computed from the expectation value of
\begin{equation}
\hat{A} \equiv e^{-\sum_i\varepsilon_i\hat{Q}_i}
\end{equation}
in the limit $\varepsilon_i\rightarrow0$ as
\begin{equation}
\ln\bigl<\hat{A}\bigr> = -\sum_i\varepsilon_i\bigl<\hat{Q}_i\bigr> + \frac{1}{2}\sum_{ij}\varepsilon_i\varepsilon_jC_{ij}+\mathcal{O}(\varepsilon^3),
\end{equation}
where $C_{ij} = \sigma^2_{Q_iQ_j}$. The Balian-V\'en\'eroni variational principle
\cite{BV(1984)}, which leads to Eq.~(\ref{Eq:sigma_XY}), is optimized for the time-evolution
of observables in the form like $\hat{A}(t)$ in the Heisenberg picture. Since higher-order terms
are neglected in the derivation, it implicitly assumes that the distribution is symmetric around the
average value. Thus, if the distribution becomes asymmetric, it may lead to unphysical results.
Our analysis confirmed that this problem arises even in a quasi-symmetric system,
$^{58}$Ni+$^{60}$Ni.

%&&&&&&&&&&&&&&&&&&&&&&&&&&&&&&&&&&&&&&&&&&&&&&&&&&
\begin{figure}[t]
   \begin{center}
   \includegraphics[width=8cm]{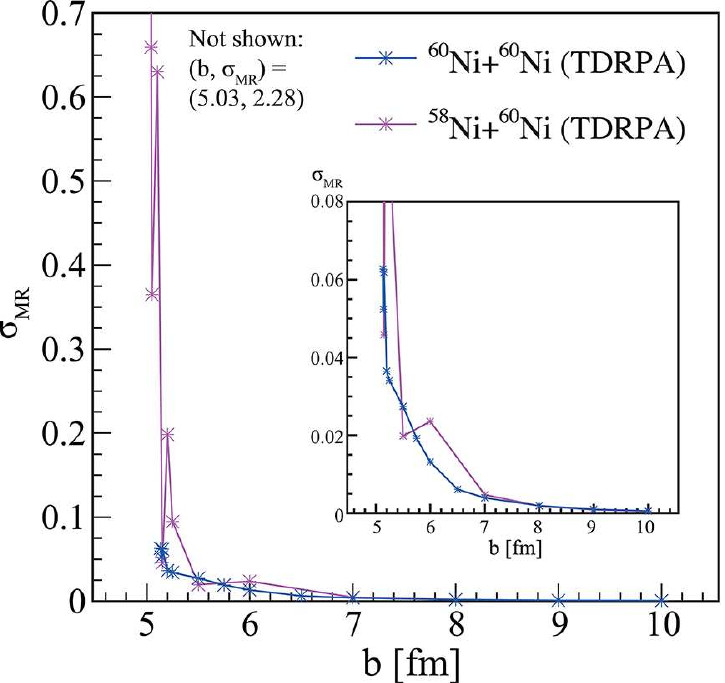}
   \end{center}\vspace{-4mm}
   \caption{
   A comparison of the $\sigma_{MR}$ values calculated using TDRPA for both $^{60}$Ni+$^{60}$Ni and $^{58}$Ni+$^{60}$Ni reactions plotted as a function of impact parameter. The inset allows a closer comparison of the two calculation sets for larger impact parameters.
   }
   \label{fig:aequiv}
\end{figure}
%&&&&&&&&&&&&&&&&&&&&&&&&&&&&&&&&&&&&&&&&&&&&&&&&&&

Figure~\ref{FIG:skewness} depicts the difference of situations encountered in (a) a symmetric
system, like $^{60}$Ni+$^{60}$Ni, and (b) in a quasi-symmetric system, like $^{58}$Ni+$^{60}$Ni.
In Fig.~\ref{FIG:skewness}~(a), a Gaussian distribution, $P(x)=\frac{1}{\sqrt{2\pi}}
\exp[-(x-\mu)^2/2\sigma^2]$, with $\mu=0.5$ and $\sigma=4$ is shown. The distribution is,
of course, symmetric by definition. In the case of symmetric reactions, the distribution must also
be symmetric and TDRPA should work well (as long as the distribution remains approximately Gaussian).
On the other hand, in Fig.~\ref{FIG:skewness}~(b), we show a skew-normal distribution with the average
value $\mu=0.5$, standard deviation $\sigma=4$ and skewness $s=0.5$, as an illustrative example.
Because of the positive skewness, the distribution is slightly leaning to the left, showing a longer tail toward
the right. As can be seen from the figure, the average value of the skewed distribution does not correspond
to the peak centroid, and the curvature around the mean is more moderate compared to the Gaussian
distribution shown in Fig.~\ref{FIG:skewness}~(a). Since TDRPA detects the curvature of the distribution
around the average value assuming the Gaussian distribution, it could provide an unphysically large value
as illustrated in the figure. One may need to extend the theoretical framework to take into account
higher-order terms, like the skewness, to correctly describe the fluctuations and correlations in
asymmetric reactions. This goes beyond the scope of the present work.  

Nevertheless, the equivalent of Fig. 3(a) in the manuscript is provided for the symmetric and asymmetric calculations in Fig. \ref{fig:aequiv}, in order to allow a visual comparison of the two results. As one can see, the $\sigma_{MR}$ values for the asymmetric reaction are unphysical for small impact parameters, but eventually converge to those for the symmetric calculations at larger impact parameters.

%&&&&&&&&&&&&&&&&&&&&&&&&&&&&&&&&&&&&&&&&&&&&&&&&&&
\begin{figure}[t]
   \begin{center}
   \includegraphics[width=7.5cm]{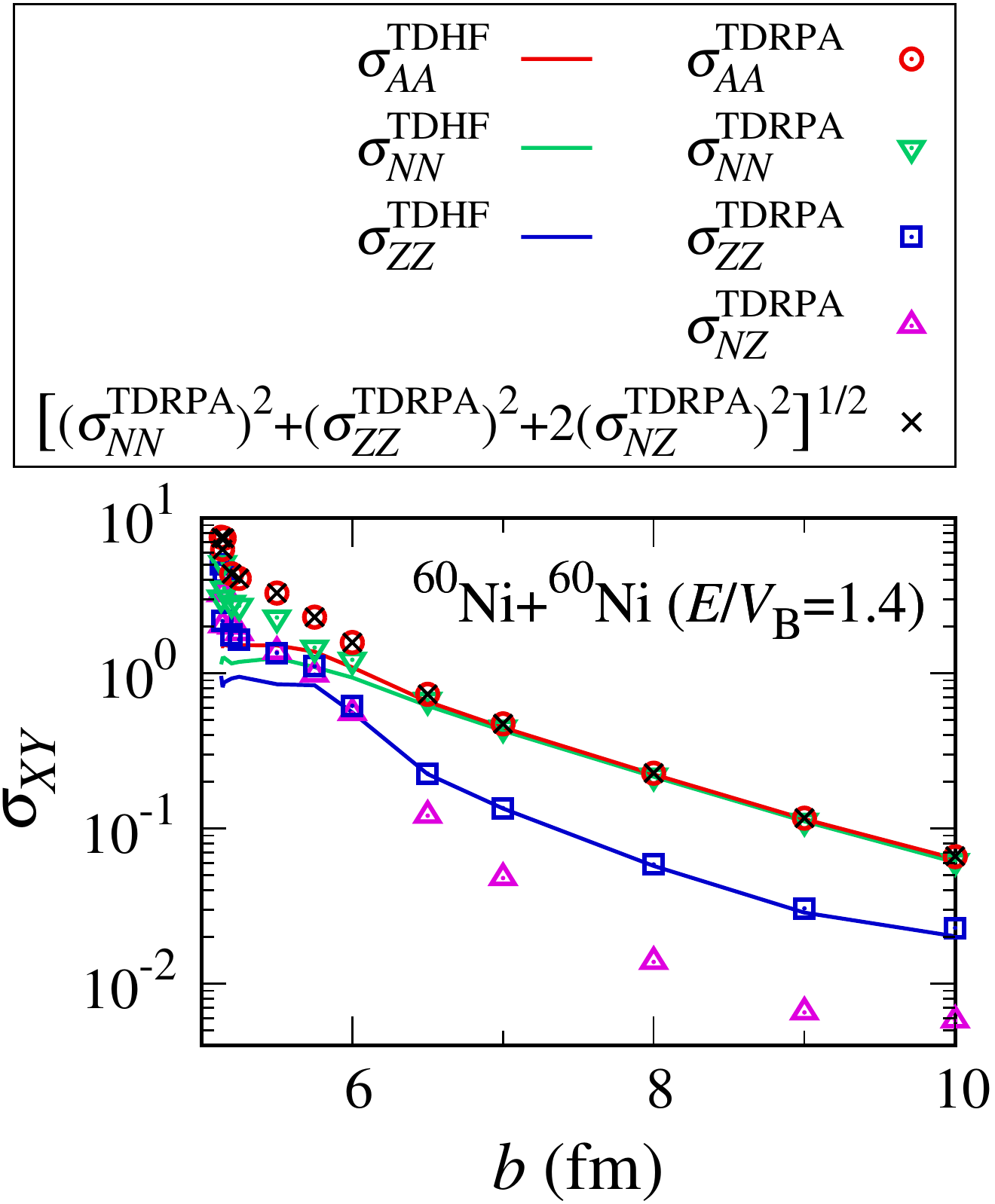}
   \end{center}\vspace{-4mm}
   \caption{
   Fluctuations and correlations $\sigma_{XY}$ for the $^{60}$Ni+$^{60}$Ni
   reaction ($E/V_B=1.40$). Fluctuations of mass, neutron, and charge numbers in TDHF
   ($\sigma_{AA}^{\rm TDHF}$, $\sigma_{NN}^{\rm TDHF}$, and $\sigma_{ZZ}^{\rm TDHF}$)
   are shown by red, green, and blue solid lines, respectively. While, those in TDRPA
   ($\sigma_{AA}^{\rm TDRPA}$, $\sigma_{NN}^{\rm TDRPA}$, and $\sigma_{ZZ}^{\rm TDRPA}$)
   are shown by red circles, green down triangles, and blue squares, respectively. Correlations between
   neutron and charge numbers in TDRPA ($\sigma_{NZ}^{\rm TDRPA}$) are shown by pink triangles.
   The mass fluctuation in TDRPA, computed by $(\sigma_{AA}^{\rm TDRPA})^2=(\sigma_{NN}^{\rm TDRPA})^2
   +(\sigma_{ZZ}^{\rm TDRPA})^2+2\,(\sigma_{NZ}^{\rm TDRPA})^2$, is also shown by black crosses.
   }
   \label{FIG:sigma_XY}
\end{figure}
%&&&&&&&&&&&&&&&&&&&&&&&&&&&&&&&&&&&&&&&&&&&&&&&&&&

\section{Fluctuations and correlations}

Let us briefly investigate fluctuations and correlations in deep-inelastic $^{60}$Ni+$^{60}$Ni
reaction at the highest incident energy examined ($E/V_B=1.40$). In Fig.~\ref{FIG:sigma_XY},
we show fluctuations of mass, neutron, and charge numbers from TDHF, $\sigma_{XX}^{\rm TDHF}$,
(solid lines) and from TDRPA, $\sigma_{XX}^{\rm TDRPA}$, (symbols), as a function of the impact
parameter. The correlation from TDRPA, $\sigma_{NZ}^{\rm TDRPA}$, is also shown by pink triangles.
Those quantities are shown in logarithmic scale.

From the figure, we find that, for reactions at relatively large impact parameters ($b\gtrsim6.5$~fm),
the fluctuations from TDHF and TDRPA are quantitatively very similar to each other. In this case,
the neutron-number fluctuation is significantly larger than that of protons, and correlations
($\sigma_{NZ}$, pink triangles) are several times smaller than the proton-number fluctuation.

As the impact parameter decreases ($b\lesssim6$~fm), we see that the fluctuations from
TDHF and TDRPA start to deviate noticeably. In this regime, the magnitude of the correlations
becomes comparable to the proton-number fluctuation, and TDRPA provides significantly
larger fluctuations as compared to the TDHF results. These results are consistent with the
observation in the $^{40}$Ca+$^{40}$Ca reaction reported in Ref.~\cite{TDRPA}.

As we computed the neutron- and proton-number fluctuations as well as the correlations
between them, we can compute the mass fluctuation from the identity, $\sigma_{AA}^2 =
\sigma_{NN}^2 + \sigma_{ZZ}^2 + 2\,\sigma_{NZ}^2$. As this expression serves an
accuracy test of the computation, the mass fluctuation computed in the latter way is also
shown in Fig.~\ref{FIG:sigma_XY} by black crosses. As can be seen from the figure, the
results coincide with those evaluated with Eq.~(\ref{Eq:sigma_XY}). Actual values are
given in Sec.~\ref{Sec:tables}.

\section{Tables of TDHF and TDRPA results}\label{Sec:tables}

Here we provide numerical results of the TDHF and TDRPA calculations
for the $^{60}$Ni+$^{60}$Ni reaction in Table~\ref{table:results}.

%===================================================
\begin{table*}[t]
\caption{
Results of the TDHF and TDRPA calculations for the $^{60}$Ni+$^{60}$Ni reaction at
$E/V_{\rm B}=1.14$, 1.27, and 1.40. From left to right, the table shows: impact parameter
$b$ in fm, corresponding angular momentum $L$ ($=b\sqrt{2\mu E}$) in units of $\hbar$,
total kinetic energy loss [TKEL ($=E-{\rm TKE}$)] in MeV, deflection angle in the center-of-mass
frame in degrees, contact time in zeptoseconds, fluctuations from TDHF $\sigma_{AA}^{\rm TDHF}$,
TDRPA $\sigma_{AA}^{\rm TDRPA}$, $\sigma_{NN}^{\rm TDRPA}$, $\sigma_{ZZ}^{\rm TDRPA}$,
and correlation from TDRPA $\sigma_{NZ}^{\rm TDRPA}$, and the mass width from TDRPA, computed
from $(\sigma_{AA}^{\rm TDRPA})^2=(\sigma_{NN}^{\rm TDRPA})^2+(\sigma_{ZZ}^{\rm TDRPA})^2
+2\,(\sigma_{NZ}^{\rm TDRPA})^2$.
}
\label{table:results}

\begin{center}
\begin{tabular}{ccccccccccc}
\hline\hline\\[-2.8mm]%_________________________________________________________
\multicolumn{11}{c}{$^{60}$Ni+$^{60}$Ni ($E/V_B=1.40$)} \\
\hline\\[-2.8mm]
$b$ (fm) & $L$ ($\hbar$) & TKEL (MeV) & $\theta_{\rm c.m.}$ (deg) & $t_{\rm contact}$ (zs) & $\sigma_{AA}^{\rm TDHF}$ & $\sigma_{AA}^{\rm TDRPA}$ & $\sigma_{NN}^{\rm TDRPA}$ & $\sigma_{ZZ}^{\rm TDRPA}$ & $\sigma_{NZ}^{\rm TDRPA}$ &
$\sigma_{AA}^{\rm TDRPA}$ \\
\hline\\[-2.9mm]
5.130 & 71.917 & 37.399 & $-$95.006 & 4.10 & 1.494 & 7.533 & 3.079 & 5.057 & 3.233 & 7.480 \\
5.140 & 72.057 & 37.088 & $-$60.531 & 3.42 & 1.493 & 6.287 & 5.142 & 2.171 & 2.040 & 6.283 \\
5.150 & 72.198 & 37.923 & $-$48.472 & 3.16 & 1.531 & 7.428 & 3.531 & 4.448 & 3.376 & 7.419 \\
5.200 & 72.899 & 37.282 & $-$24.952 & 2.73 & 1.482 & 4.378 & 2.841 & 1.769 & 1.996 & 4.378 \\
5.250 & 73.599 & 36.231 & $-$14.245 & 2.51 & 1.518 & 4.089 & 2.723 & 1.651 & 1.813 & 4.089 \\
5.500 & 77.104 & 35.071 & 5.512    & 2.14 & 1.514 & 3.292 & 2.288 & 1.352 & 1.373 & 3.292 \\
5.750 & 80.609 & 35.920 & 44.395  & 1.37 & 1.378 & 2.301 & 1.464 & 1.108 & 0.981 & 2.301 \\
6.000 & 84.114 & 14.971 & 59.472 & 0.94 & 1.093 & 1.578 & 1.222 & 0.617 & 0.555 & 1.578 \\
6.250 & 87.618 & 4.904 & 62.702   & 0.76 & 0.821 & 0.974 & 0.866 & 0.320 & 0.220 & 0.974 \\
6.500 & 91.123 & 2.608 & 62.481  & 0.68 & 0.656 & 0.732 & 0.675 & 0.226 & 0.121 & 0.732 \\
7.000 & 98.133 & 1.274 & 60.178 & 0.57 & 0.447 & 0.472 & 0.448 & 0.135 & 0.048 & 0.473 \\
8.000 & 112.152 & 0.624 & 54.536 & 0.38 & 0.223 & 0.227 & 0.219 & 0.059 & 0.014 & 0.228 \\
9.000 & 126.170 & 0.432 & 49.408 & 0.00 & 0.115 & 0.116 & 0.112 & 0.031 & 0.007 & 0.117 \\
10.000 & 140.189 & 0.342 & 45.040 & 0.00 & 0.064 & 0.066 & 0.062 & 0.023 & 0.006 & 0.066 \\
\hline\\[-2.8mm]
\multicolumn{11}{c}{$^{60}$Ni+$^{60}$Ni ($E/V_B=1.27$)} \\
\hline\\[-2.8mm]
$b$ (fm) & $L$ ($\hbar$) & TKEL (MeV) & $\theta_{\rm c.m.}$ (deg) & $t_{\rm contact}$ (zs) & $\sigma_{AA}^{\rm TDHF}$ & $\sigma_{AA}^{\rm TDRPA}$ & $\sigma_{NN}^{\rm TDRPA}$ & $\sigma_{ZZ}^{\rm TDRPA}$ & $\sigma_{NZ}^{\rm TDRPA}$ &
$\sigma_{AA}^{\rm TDRPA}$ \\
\hline\\[-2.9mm]
4.790 & 63.958 & 28.412 & $-$33.574 & 3.52 & 1.446 & 5.585 & 7.733 & 9.338 & 7.613 & 16.214 \\
4.800 & 64.092 & 26.636 & $-$9.072 & 2.97 & 1.439 & 6.637 & 3.900 & 3.036 & 3.128 & 6.632 \\
4.850 & 64.759 & 24.999 & 31.020 & 2.10 & 1.397 & 3.104 & 2.261 & 1.094 & 1.289 & 3.103 \\
4.900 & 65.427 & 28.466 & 43.179 & 1.80 & 1.391 & 2.737 & 1.768 & 1.258 & 1.179 & 2.737 \\
4.950 & 66.094 & 30.291 & 51.123 & 1.62 & 1.327 & 2.393 & 1.553 & 1.111 & 1.019 & 2.393 \\
5.000 & 66.762 & 28.301 & 57.338 & 1.47 & 1.268 & 1.986 & 1.336 & 0.918 & 0.811 & 1.986 \\
5.250 & 70.100 & 12.010 & 72.917 & 1.00 & 1.025 & 1.456 & 1.158 & 0.536 & 0.496 & 1.456 \\
5.500 & 73.438 & 3.904 & 75.313 & 0.81 & 0.779 & 0.908 & 0.823 & 0.265 & 0.196 & 0.908 \\
5.750 & 76.776 & 2.215 & 74.451 & 0.73 & 0.636 & 0.699 & 0.654 & 0.191 & 0.113 & 0.699 \\
6.000 & 80.114 & 1.537 & 72.944 & 0.68 & 0.531 & 0.567 & 0.538 & 0.149 & 0.072 & 0.568 \\
7.000 & 93.467 & 0.694 & 65.881 & 0.47 & 0.274 & 0.280 & 0.272 & 0.066 & 0.018 & 0.281 \\
8.000 & 106.819 & 0.453 & 59.376 & 0.19 & 0.145 & 0.146 & 0.142 & 0.034 & 0.009 & 0.147 \\
9.000 & 120.172 & 0.344 & 53.832 & 0.00 & 0.079 & 0.080 & 0.077 & 0.023 & 0.007 & 0.081 \\
10.000 & 133.524 & 0.286 & 49.147 & 0.00 & 0.048 & 0.050 & 0.046 & 0.020 & 0.007 & 0.051 \\
\hline\\[-2.8mm]
\multicolumn{11}{c}{$^{60}$Ni+$^{60}$Ni ($E/V_B=1.14$)} \\
\hline\\[-2.8mm]
$b$ (fm) & $L$ ($\hbar$) & TKEL (MeV) & $\theta_{\rm c.m.}$ (deg) & $t_{\rm contact}$ (zs) & $\sigma_{AA}^{\rm TDHF}$ & $\sigma_{AA}^{\rm TDRPA}$ & $\sigma_{NN}^{\rm TDRPA}$ & $\sigma_{ZZ}^{\rm TDRPA}$ & $\sigma_{NZ}^{\rm TDRPA}$ &
$\sigma_{AA}^{\rm TDRPA}$ \\
\hline\\[-2.9mm]
3.740 & 47.313 & 17.896 & 45.713 & 2.68 & 1.253 & 3.043 & 2.040 & 1.816 & 0.924 & 3.028 \\
3.750 & 47.439 & 17.691 & 56.969 & 2.37 & 1.260 & 3.683 & 2.779 & 1.182 & 1.483 & 3.677 \\
3.800 & 48.072 & 19.416 & 73.002 & 1.83 & 1.186 & 2.034 & 1.503 & 0.965 & 0.689 & 2.034 \\
3.850 & 48.704 & 18.950 & 80.793 & 1.58 & 1.109 & 1.682 & 1.138 & 0.783 & 0.678 & 1.682 \\
3.900 & 49.337 & 18.937 & 84.244 & 1.46 & 1.060 & 1.548 & 1.058 & 0.709 & 0.623 & 1.548 \\
3.950 & 49.969 & 18.918 & 86.766 & 1.38 & 1.056 & 1.486 & 1.037 & 0.673 & 0.582 & 1.486 \\
4.000 & 50.602 & 16.850 & 89.906 & 1.28 & 1.065 & 1.675 & 1.145 & 0.744 & 0.686 & 1.675 \\
4.250 & 53.765 & 4.230 & 95.617 & 0.93 & 0.812 & 0.999 & 0.914 & 0.247 & 0.225 & 0.999 \\
4.375 & 55.346 & 2.962 & 95.085 & 0.87 & 0.733 & 0.854 & 0.799 & 0.196 & 0.162 & 0.854 \\
4.500 & 56.927 & 2.291 & 94.204 & 0.83 & 0.670 & 0.755 & 0.715 & 0.167 & 0.126 & 0.756 \\
4.750 & 60.090 & 1.598 & 92.033 & 0.77 & 0.570 & 0.618 & 0.593 & 0.132 & 0.084 & 0.618 \\
5.000 & 63.252 & 1.236 & 89.664 & 0.72 & 0.491 & 0.520 & 0.502 & 0.108 & 0.058 & 0.520 \\
6.000 & 75.903 & 0.650 & 80.228 & 0.51 & 0.276 & 0.280 & 0.275 & 0.055 & 0.019 & 0.281 \\
7.000 & 88.553 & 0.435 & 71.938 & 0.20 & 0.153 & 0.155 & 0.152 & 0.031 & 0.009 & 0.156 \\
8.000 & 101.204 & 0.333 & 64.931 & 0.00 & 0.086 & 0.088 & 0.085 & 0.022 & 0.006 & 0.088 \\
9.000 & 113.854 & 0.268 & 59.020 & 0.00 & 0.052 & 0.053 & 0.050 & 0.019 & 0.006 & 0.054 \\
10.000 & 126.505 & 0.227 & 54.013 & 0.00 & 0.036 & 0.038 & 0.035 & 0.017 & 0.006 & 0.040 \\
\hline\hline%_________________________________________________________
\end{tabular}
\end{center}
\end{table*}
%===================================================

% ============================================================================
% ================================ SUPPLEMENT ==================================
% ============================================================================

\end{document}